# Slowing Down of Charged Particles in Dusty Plasmas with Power-law Kappa-distributions

Jiulin Du [1*], Ran Guo [2], Zhipeng Liu [3] and Songtao Du [4]

[1] *Department of Physics, School of Science, Tianjin University, Tianjin 300072, China*

[2] *School of Science, Civil Aviation University of China, Tianjin 300300, China*

[3] *School of Science, Tianjin Chengjian University, Tianjin 300384, China*

[4] *College of Electronic Information and Automation, Civil Aviation University of China, Tianjin 300300, China*

**Keywords:** Slowing down, Kappa-distributions, Dusty plasma, Fokker-Planck collision theory

**Abstract** We study slowing down of a particle beam passing through the dusty plasma with power-law $\kappa$-distributions. Three plasma components, electrons, ions and dust particles, can have a different $\kappa$-parameter. The deceleration factor and slowing down time are derived and expressed by a hyper-geometric $\kappa$-function. Numerically we study slowing down property of an electron beam in the $\kappa$-distributed dusty plasma. We show that the slowing down in the plasma depends strongly on the $\kappa$-parameters of plasma components, and dust particles play a dominant role in the deceleration effects. We also show dependence of the slowing down on mass and charge of a dust particle in the plasma.

## 1 Introduction

Dusty plasmas are ubiquitous in astrophysical, space and terrestrial environments, such as the interstellar clouds, the circumstellar clouds, the interplanetary space, the comets, the planetary rings, the Earth's atmosphere, and the lower ionosphere etc. They can also exist in laboratory plasma environments. Dusty plasma consists of three components: electrons, ions and dust particles of micron- or/and submicron-sized particulates. Dust particles can get charges through a variety of interactions with electrons and ions. These interactions may also take place between the dust particles and they can make a response to an external perturbation.

When a charged particle beam flies into the plasma, it will be decelerated due to the resistance of each component of the plasma. This phenomenon is called slowing down. The slowing down of a charged particle beam can be studied by using some appropriate statistical description, such as the Fokker-Planck collision theory for fully ionized plasma [1]. The velocity distribution of a particle beam (as a test particle), when it flies into the plasma, will change due to its collisions with plasma component particles. If plasma is assumed to be in thermodynamic equilibrium state and follow a Maxwellian distribution, the slowing down phenomenon can be discussed by an error function on the mean velocity of the particle beam [2]. But if the plasma is nonequilibrium and the velocity distributions are non-Maxwellian or power-law ones, properties of the slowing down are not known yet.

The power-law distributions usually include the $q$-distributions in nonextensive

---

statistical mechanics [3] as well as the kappa-distributions in astrophysical and space plasmas. The two types of distributions now can be thought to be equivalent to each other in plasma physics, and the kappa-distributions can be studied within the framework of nonextensive statistics.

Non-Maxwellian distributions can be observed commonly both in astrophysical, space plasmas and in laboratory plasmas. In fact, spacecraft measurements of plasma velocity distributions, both in the solar wind and in planetary magnetospheres and magnetosheaths, have revealed that non-Maxwellian distributions are quite common. In many situations, the distributions appear reasonably Maxwellian at low energies but have a "suprathermal" power-law tail at high energies, and the power-law tail can be well modeled by the so-called kappa-distributions. The kappa-distributions ($\kappa$-distributions) are power-law distributions and the kappa-distribution function for the velocity can be expressed [4] as follows:

$$f_\kappa(\mathbf{v}) = B_\kappa \left(1 + \frac{1}{2\kappa+3}\frac{m\mathrm{v}^2}{kT}\right)^{-(\kappa+1)}, \kappa > \frac{3}{2}, \tag{1}$$

where $\kappa$ is a positive parameter, $k$ is Boltzmaan constant, $T$ is temperature, $m$ is mass of a particle, and $B_\kappa$ is the normalized coefficient given by

$$B_\kappa = \left[\frac{\pi kT}{m}(2\kappa-3)\right]^{-3/2} \frac{\Gamma(\kappa+1)}{\Gamma(\kappa-\frac{1}{2})}.$$

The velocity $\kappa$-distribution can return to a Maxwellian velocity distribution if the $\kappa$-parameter takes the limit $\kappa \to \infty$. The $\kappa$-parameter is a very important quantity to characterize the complex plasma with power-law velocity distributions.

In the past many years, many attempts have been made to explore the physical nature of the $q$-parameter and the $\kappa$-parameter in dynamical systems. For the nonequilibrium plasma with Coulomb long-range interactions, the equation of the $q$-parameter in nonextensive statistics was first derived in 2004 [5] as

$$k\nabla T = e(1-q)\nabla\varphi_c, \tag{2}$$

where $e$ is electron charge, and $\varphi_c$ is Coulomb potential. The $q$-parameter was therefore given a clear physical meaning for $q\neq 1$. And further, if a nonequilibrium plasma has magnetic field, convection and rotation, the $q$-parameter or/and the $\kappa$-parameter can be determined by the following most general equation [6],

$$(\kappa-\frac{3}{2})k\nabla T = e[-\nabla\varphi_c + c^{-1}\mathbf{u}\times\mathbf{B}] - m(\omega^2\mathbf{R} + 2\mathbf{u}\times\boldsymbol{\omega}), \tag{3}$$

where $c$ is light speed, $\mathbf{u}$ is fluid motion velocity and $\mathbf{B}$ is magnetic induction intensity. Now we have well known that the equation for the $q$-parameter or/and the $\kappa$-parameter represents the nonequilibrium nature of complex plasma with the power-law distributions, having electromagnetic field, convection and rotation. The rotation effect contains two terms, $m\omega^2\mathbf{R} + 2m\mathbf{u}\times\boldsymbol{\omega}$. The first term $m\omega^2\mathbf{R}$ is the contribution due to the inertial centrifugal force. When the angle velocity $\omega$ varies from the equator to the poles, differential rotation exists depending on the angle velocity $\omega$ and the vertical distance $\mathbf{R}$ between the particle and the rotation axis. The second term $2m\mathbf{u}\times\mathbf{\omega}$ is the contribution due to Coriolis force, depending on the angle velocity $\mathbf{\omega}$ and motion velocity $\mathbf{u}$ of the fluid. For the case without magnetic field and rotation, we just need to take $\mathbf{B}=0$ and $\omega=0$ in Eq.(3).

The kappa-distribution was drawn first by Vasyliunas in 1968 as an empirical function to model the velocity distribution of observed electrons within the plasma sheet of the magnetosphere [7]. Later, in solar corona, $\kappa$-like power-law distributions were proposed to arise from strong nonequilibrium thermodynamic gradients, Fermi acceleration at upwelling convection-zone waves or shocks, and electron-ion runaway in a Dreicer-order electric field (see Ref.[4] and the references therein). In solar wind plasmas, a log-kappa distribution was introduced so as to replace the log-normal distribution [8, 9]. Up to now, the $\kappa$-distributions have aroused great concern for its new characteristics and interesting applications found in many fields of astrophysics, space plasmas and laboratory plasmas, for example, $\kappa$-distribution in a superthermal radiation field plasma [10], a solar wind kinetic model based on the $\kappa$-distributions for electrons and protons escaping out of solar corona [11], Landau damping, ion acoustic waves and dust acoustic waves in space plasmas [12-15], fluctuations in magnetized plasmas with $\kappa$-distributions [16], some transport coefficients in $\kappa$-distributed plasmas [17,18], and properties of the $\kappa$-distributions in space plasmas [19] etc. In particular, with the development of nonextensive statistics, when it is recognized that this new statistical theory can be used as a theoretical basis for the study of power-law distributed plasmas, the investigations of $q$-distributed plasmas are becoming increasingly widespread across both astrophysics and space science with an exponential growth rate of relevant publications (see the list of publications on plasma physics in Ref.[3]), which include a wide variety of waves and instabilities both in different electron-ion plasmas and dusty plasmas, solar wind and properties of other plasmas etc [20-33].

In this work, we study the Fokker-Planck theory for the slowing down phenomena of a particle beam passing through the dusty plasma with power-law $\kappa$-distributions. In section 2, we discuss basic FP theory of a test particle in the $\kappa$-distributed dusty plasma. In section 3, we derive the deceleration factor (the velocity moment equation of the test particle) and the slowing down time. In section 4, we take an electron beam as an example to make numerical analyses to show the dependence of the slowing down property on different $\kappa$-parameters. In section 5, we present our conclusion.

## 2 The Fokker-Planck theory for the $\kappa$–distributed dusty plasmas

The study of transport properties in nonequilibrium plasma is usually based on a collision model which can simulate the interactions between the particles and provide a mathematical simplification. According to different plasma states, we can find an appropriate statistical theory to construct the collision model. In fully ionized plasma, due to the long-range Coulomb interactions of charged particles, it is assumed that a rapid large angle deflection of a charged particle by the particle-particle collision is the result of continuous small angle scattering from the long distance particles in background plasma. In other words, when a charged particle moves along its path in plasma, it has undergone many small angles of Coulomb scattering. Based on this idea, the collision model known as Fokker-Planck (FP) collision theory is discussed, giving the following FP equation [1,2],

$$\frac{\partial f_T}{\partial t} = \sum_\alpha \frac{4\pi n_\alpha q_T^2 q_\alpha^2}{m_T^2} \left[ -\frac{\partial}{\partial \mathbf{v}} \cdot \left( f_T \frac{\partial h_\alpha}{\partial \mathbf{v}} \right) + \frac{1}{2} \frac{\partial}{\partial \mathbf{v}} \frac{\partial}{\partial \mathbf{v}} : \left( f_T \frac{\partial^2 g_\alpha}{\partial \mathbf{v} \partial \mathbf{v}} \right) \right], \tag{4}$$

where $f_T \equiv f_T(\mathbf{v}, t)$ is a velocity distribution function of test particles (or the particle beam), $q_T$ and $m_T$ are respectively charge and mass of the test particle, $q_\alpha$ and $n_\alpha$ are respectively charge and number density of the $\alpha$th component particles in the plasma. The functions $h_\alpha$ and $g_\alpha$ are expressed respectively by

$$h_\alpha(\mathbf{v}) = \frac{m_T}{\mu_\alpha} \int \frac{f_\alpha(\mathbf{v}')}{|\mathbf{v} - \mathbf{v}'|} d\mathbf{v}' \tag{5}$$

and

$$g_\alpha(\mathbf{v}) = \int f_\alpha(\mathbf{v}') |\mathbf{v} - \mathbf{v}'| d\mathbf{v}', \tag{6}$$

where the reduced mass of the $\alpha$th component particle is $\mu_\alpha = m_T m_\alpha / (m_T + m_\alpha)$. The function $f_\alpha(\mathbf{v})$ is a stationary velocity distribution of the $\alpha$th component particles in the plasma. On the right-hand side of Eq.(4), the first term is a friction term, which describes the slowing down of the velocity associated with the distribution $f_T$; while the second term is a diffusion term, which describes the spreading out (i.e., diffusive broadening) of the velocity distribution described by $f_T$. The FP equation thus gives the rate of change of the distribution function of a test particle due to collisions with all the $\alpha$ components of particles in the plasma. It is clear that the FP equation describes the particle-particle interaction between a test particle and charged particles in background plasma, where change in energy of the test particle can be discussed by both the friction and diffusion term (diffusive acceleration and friction deceleration) [34].

In the past many situations, normally it has been assumed that the velocity distributions of the plasma particles obey a Maxwellian distribution, which is actually equivalent to the plasma being in a thermal equilibrium state. Now we consider the nonequilibrium dusty plasma with the power-law $\kappa$-distributions. If we let the subscript $\alpha = e, i,$ and $d$ stands for electrons, ions, and dust particles respectively, and their densities are denoted by $n_e$, $n_i$, and $n_d$ respectively, the charge neutrality condition is written as

$$\varepsilon_d Z_d n_d + Z_i n_i - n_e = 0,$$

where $Z_d$ and $Z_i$ are the charge number of the dust particle and the ion, respectively; $\varepsilon_d = +1$ (or $-1$) is for positively (or negatively) charged dust particle. In the dusty plasma if the three components have different temperature $T_\alpha$ and different $\kappa$-parameter $\kappa_\alpha$, then according to Eq.(1) the velocity $\kappa$-distribution function for the $\alpha$th component particles can be expressed as

$$f_{\kappa,\alpha}(\mathbf{v}) = B_{\kappa,\alpha} \left( 1 + \frac{1}{2\kappa_\alpha - 3} \frac{m_\alpha \mathbf{v}^2}{kT_\alpha} \right)^{-(\kappa_\alpha + 1)}, \quad \kappa_\alpha > \frac{3}{2}, \tag{7}$$

and correspondingly, the normalization coefficient $B_{\kappa,\alpha}$ becomes

$$B_{\kappa,\alpha} = \left[\frac{\pi k T_\alpha}{m_\alpha}(2\kappa_\alpha - 3)\right]^{-3/2} \frac{\Gamma(\kappa_\alpha + 1)}{\Gamma(\kappa_\alpha - \frac{1}{2})}.$$

In theory, the *q*-distribution in nonextensive kinetics is a stationary solution of the generalized Boltzmann equation at a *q*-collision equilibrium [35], the *q*- or/and *κ*-distribution functions are also stationary solutions of the FP equation when a generalized fluctuation-dissipation relation between the friction and diffusion coefficient is satisfied in a complex system [36], therefore the *q*- or/and *κ*-distribution is a physical state that is still dominated by collisions. In the solar flare plasma, the *κ*-distribution function was determined by a balance between diffusive acceleration and collisions [37]. Although some collisionless space plasmas were often modeled by the *κ*-distributions, it has been known that the *κ*-distributions have significant impact on the resulting optically thin spectra arising from collisionally dominated astrophysical plasmas, and the relevant collision process rate coefficients are provided for *κ* values of the *κ*-distributed plasmas [38,39]. Here based on the FP theory, we study the slowing down of a test particle in the *κ*-distributed dusty plasma.

## 3 The slow down of a particle beam in the *κ*-distributed dusty plasma

In the dusty plasma, we consider a test particle with density $n_T$, mass $m_T$, charge $q_T$ and the velocity distribution function at time *t*=0,

$$f_T(\mathbf{v}, t=0) = \delta(\mathbf{v} - \mathbf{v}_0). \tag{8}$$

The velocity moment (mean velocity) of the test particle is defined as

$$\mathbf{U} = \int \mathbf{v}\, f_T(\mathbf{v},t)d\mathbf{v}. \tag{9}$$

Therefore, using the Fokker-Planck equation (4), we can obtain the velocity moment equation (i.e. the deceleration factor) by integration by parts on the right-hand side terms of Eq.(4) [1], i.e.,

$$-\int \mathbf{v} \frac{\partial}{\partial \mathbf{v}} \cdot \left( f_T \frac{\partial h_{\kappa,\alpha}}{\partial \mathbf{v}} \right) d\mathbf{v} = \int f_T \frac{\partial h_{\kappa,\alpha}}{\partial \mathbf{v}} d\mathbf{v}, \text{ and}$$

$$\int \mathbf{v} \frac{\partial}{\partial \mathbf{v}} \frac{\partial}{\partial \mathbf{v}} : \left( f_T \frac{\partial^2 g_{\kappa,\alpha}}{\partial \mathbf{v} \partial \mathbf{v}} \right) d\mathbf{v} = -\int \frac{\partial}{\partial \mathbf{v}} \cdot \left( f_T \frac{\partial^2 g_{\kappa,\alpha}}{\partial \mathbf{v} \partial \mathbf{v}} \right) d\mathbf{v} = 0.$$

The velocity moment equation is therefore determined by

$$\frac{\partial \mathbf{U}}{\partial t} = \frac{4\pi q_T^2 \ln \Lambda}{m_T^2} \sum_\alpha q_\alpha^2 n_\alpha \int d\mathbf{v}\, f_T(\mathbf{v},t) \frac{\partial h_{\kappa,\alpha}(\mathbf{v})}{\partial \mathbf{v}}, \tag{10}$$

where $n_\alpha$ is density, $q_\alpha$ is charge, $\ln\Lambda$ is the scattering factor between the test particle and the plasma. Substituting the distribution function (8) of the test particle at time *t*=0 into Eq.(10), we can study the slowing down of the particle beam at time *t*=0 in the plasma by the following moment equation,

$$\left.\frac{\partial \mathbf{U}}{\partial t}\right|_{t=0} = \frac{4\pi q_T^2 \ln \Lambda}{m_T^2} \sum_\alpha q_\alpha^2 n_\alpha \frac{\partial}{\partial \mathbf{v}_0} h_{\kappa,\alpha}(\mathbf{v}_0)$$

$$= \frac{4\pi q_T^2 \ln \Lambda}{m_T^2} \sum_\alpha q_\alpha^2 n_\alpha \frac{\partial}{\partial \mathbf{U}} h_{\kappa,\alpha}(\mathbf{U}). \tag{11}$$

where $\mathbf{U}$ is used as the mean velocity $\mathbf{v}_0$ at $t=0$.

In our case, the velocity distribution functions of dusty plasma components are assumed to be the power-law $\kappa$-distributions. Therefore, according to Eq.(5) the function $h_{\kappa,\alpha}(\mathbf{v})$ can be written by

$$h_{\kappa,\alpha}(\mathbf{v}) = \frac{m_T}{\mu_\alpha} \int \frac{f_{\kappa,\alpha}(\mathbf{v}')}{|\mathbf{v}-\mathbf{v}'|} d\mathbf{v}'$$

$$= \frac{m_T}{\mu_\alpha} B_{\kappa,\alpha} \int \frac{d\mathbf{v}'}{|\mathbf{v}-\mathbf{v}'|} \left(1 + \frac{1}{2\kappa_\alpha - 3} \frac{m_\alpha v'^2}{kT_\alpha}\right)^{-(\kappa_\alpha+1)}. \tag{12}$$

In order to calculate this function, we use a variable substitution $\mathbf{w} = \mathbf{v}'-\mathbf{v}$, where the direction of $\mathbf{v}$ defines the axis of a spherical polar coordinate system and $\varphi$ is the angle between $\mathbf{w}$ and $\mathbf{v}$. Then the equation (12) is rewritten as

$$h_{\kappa,\alpha}(\mathbf{v}) = \frac{\pi m_T}{\mu_\alpha} \frac{kT_\alpha}{m_\alpha v} \left(2 - \frac{3}{\kappa_\alpha}\right) B_{\kappa,\alpha} \left\{ \int_0^\infty dw \left[1 + \frac{m_\alpha(w\text{-}v)^2}{(2\kappa_\alpha-3)kT_\alpha}\right]^{-\kappa_\alpha} \right.$$

$$\left. - \int_0^\infty dw \left[1 + \frac{m_\alpha(w+v)^2}{(2\kappa_\alpha-3)kT_\alpha}\right]^{-\kappa_\alpha} \right\}. \tag{13}$$

On the right-hand side in Eq.(13), we use symbol $A_\alpha \equiv m_\alpha/[(2\kappa_\alpha-3)kT_\alpha]$ and then we let $y = \sqrt{A_\alpha}(w\text{-}v)$ in the first integral and let $y = \sqrt{A_\alpha}(w+v)$ in the second integral. This equation (13) therefore becomes the following form (see Appendix),

$$h_{\kappa,\alpha}(\mathbf{v}) = \frac{2\pi m_T}{\mu_\alpha v} \left[\frac{kT_\alpha}{m_\alpha}(2\kappa_\alpha - 3)\right]^{3/2} \frac{B_{\kappa,\alpha}}{\kappa_\alpha} \int_0^{\sqrt{A_\alpha}v} dy(1+y^2)^{-\kappa_\alpha}. \tag{14}$$

Further making the variable substitution, $y^2 = A_\alpha v^2 x$, in Eq.(14), we have that

$$h_{\kappa,\alpha}(\mathbf{v}) = \frac{\pi m_T kT_\alpha}{\mu_\alpha m_\alpha} \frac{2\kappa_\alpha - 3}{\kappa_\alpha} B_{\kappa,\alpha} \int_0^1 x^{-\frac{1}{2}}(1+A_\alpha v^2 x)^{-\kappa_\alpha} dx. \tag{15}$$

Now we use the integral representation of a hyper-geometric function defined [40] by

$$_2F_1(a,b;\,c;\,z) = \frac{\Gamma(c)}{\Gamma(a)\Gamma(c-a)} \int_0^1 x^{a-1}(1-x)^{c-a-1}(1-zx)^{-b}dx, \quad |z|<1, \tag{16}$$

and then equation (16) can be written as

$$h_{\kappa,\alpha}(\mathbf{v}) = \frac{2\pi\, m_T\, kT_\alpha}{\mu_\alpha m_\alpha} \frac{2\kappa_\alpha - 3}{\kappa_\alpha} B_{\kappa,\alpha}\, _2F_1\left[\frac{1}{2},\, \kappa_\alpha;\, \frac{3}{2};\, -A_\alpha v^2\right]. \tag{17}$$

Now merging $B_{\kappa,\alpha}$ into Eq.(17) and then substituting Eq.(17) into Eq.(11), the velocity moment equation of the test particle can be expressed by a hyper-geometric function. Namely, the deceleration factor (11) can be written as

$$\left.\frac{\partial \mathbf{U}}{\partial t}\right|_{t=0} = 8\sqrt{\pi}\, q_T^2\, m_T^{-2} \ln\Lambda \sum_\alpha C_{\kappa,\alpha} \frac{\partial}{\partial \mathbf{U}}\, {}_2F_1\left(\frac{1}{2}, \kappa_\alpha; \frac{3}{2}; -A_\alpha \mathbf{U}^2\right)$$

$$= \frac{16}{3}\sqrt{\pi}\, q_T^2 m_T^{-2} \mathbf{U} \ln\Lambda \sum_\alpha -D_{\kappa,\alpha}\, {}_2F_1\left(\frac{3}{2}, \kappa_\alpha+1; \frac{5}{2}; -A_\alpha \mathbf{U}^2\right), \quad (18)$$

where $C_{\kappa,\alpha}$ and $D_{\kappa,\alpha}$ are respectively

$$C_{\kappa,\alpha} = q_\alpha^2 n_\alpha \left(1+\frac{m_T}{m_\alpha}\right)\left[\frac{kT_\alpha}{m_\alpha}(2\kappa_\alpha-3)\right]^{-1/2} \frac{\Gamma(\kappa_\alpha)}{\Gamma(\kappa_\alpha-1/2)}, \text{ and}$$

$$D_{\kappa,\alpha} = q_\alpha^2 n_\alpha \left(1+\frac{m_T}{m_\alpha}\right)\left[\frac{kT_\alpha}{m_\alpha}(2\kappa_\alpha-3)\right]^{-3/2} \frac{\Gamma(\kappa_\alpha+1)}{\Gamma(\kappa_\alpha-\frac{1}{2})}. \quad (19)$$

On the right-hand side of Eq.(18), the three terms represent the resistances to the test particles due to electrons, ions and dust particles, respectively, in the plasma. By using Eq.(18) we can study the slowing down property of a particle beam passing through the nonequilibrium dusty plasma with power-law $\kappa$-distributions. The particle beam can be considered as electrons, ions, or dust particles.

On the other hand, the slowing down time is defined [2] by

$$\tau_s = -\mathbf{U}\left(\left.\frac{\partial \mathbf{U}}{\partial t}\right|_{t=0}\right)^{-1}. \quad (20)$$

And thus in our case, we have that

$$\tau_s = \frac{3 m_T^2}{16\sqrt{\pi}\, q_T^2 \ln\Lambda}\left[\sum_\alpha D_{\kappa,\alpha}\, {}_2F_1\left(\frac{3}{2}, \kappa_\alpha+1; \frac{5}{2}; -A_\alpha \mathbf{U}^2\right)\right]^{-1}. \quad (21)$$

Just as we expected, when we take $\kappa\to\infty$, Eq.(18) and Eq.(21) are both reduced to the equations in the case of thermal equilibrium plasma with a Maxwellian velocity distribution. Namely [2],

$$\left.\frac{\partial \mathbf{U}}{\partial t}\right|_{t=0} = 8\sqrt{\pi}\, q_T^2 m_T^{-2} \mathbf{U} \ln\Lambda \sum_\alpha -D_{\infty,\alpha}\int_0^1 x^{\frac{1}{2}} e^{-\frac{m_\alpha \mathbf{U}^2}{2kT_\alpha}x} dx, \quad (22)$$

with $D_{\infty,\alpha} = q_\alpha^2 n_\alpha (1+m_T/m_\alpha)[2kT_\alpha/m_\alpha]^{-3/2}$, and

$$\tau_s = \frac{m_T^2}{8\sqrt{\pi}\, q_T^2 \ln\Lambda}\left[\sum_\alpha -D_{\infty,\alpha}\int_0^1 x^{\frac{1}{2}} e^{-\frac{m_\alpha \mathbf{U}^2}{2kT_\alpha}x} dx\right]^{-1}. \quad (23)$$

As we well known, they are expressed by an error function.

**4 Numerical analyses for different $\kappa$-parameters**

In order to study dependence of the slowing down property on different $\kappa$-parameters and then to understand the deceleration effects of a charged particle beam passing through the dusty plasma with power-law $\kappa$-distributions, in this section we make the numerical analyses of the deceleration factor. To analyze the role of each plasma component in the deceleration factor, we can write Eq.(18) as three parts: the electron term, the ion term and the dust term, namely,

$$\left.\frac{\partial \mathbf{U}}{\partial t}\right|_{t=0} = \left(\frac{\partial \mathbf{U}}{\partial t}\right)_e + \left(\frac{\partial \mathbf{U}}{\partial t}\right)_i + \left(\frac{\partial \mathbf{U}}{\partial t}\right)_d, \qquad (24)$$

where the electron term is

$$\left(\frac{\partial \mathbf{U}}{\partial t}\right)_e = -\frac{16\ln\Lambda}{3}\sqrt{\pi}q_T^2 m_T^{-2} D_{\kappa,e}\mathbf{U}\ _2F_1\left(\frac{3}{2},\kappa_e+1;\frac{5}{2};-A_e\mathbf{U}^2\right), \qquad (25)$$

the ion term is

$$\left(\frac{\partial \mathbf{U}}{\partial t}\right)_i = -\frac{16\ln\Lambda}{3}\sqrt{\pi}q_T^2 m_T^{-2} D_{\kappa,i}\mathbf{U}\ _2F_1\left(\frac{3}{2},\kappa_i+1;\frac{5}{2};-A_i\mathbf{U}^2\right), \qquad (26)$$

and the dust term is

$$\left(\frac{\partial \mathbf{U}}{\partial t}\right)_d = -\frac{16\ln\Lambda}{3}\sqrt{\pi}q_T^2 m_T^{-2} D_{\kappa,d}\mathbf{U}\ _2F_1\left(\frac{3}{2},\kappa_d+1;\frac{5}{2};-A_d\mathbf{U}^2\right). \qquad (27)$$

They give the deceleration factors respectively caused by electrons, ions and/or dust particles in the background plasma.

From Eqs.(25)-(27) we can see that the slowing down effects depend on the mean velocity, charge and mass of the beam's particle, but also depend on the physical quantities of the dusty plasma, such as temperature, mass, density and charge of each component particle. The three parts in Eq.(24), the electron term, the ion term and the dust particle term, all can be described by a hypergeometric function about the $\kappa$-parameter and the beam's mean velocity. Although these physical quantities in space plasmas are quite uncertain, the nominal values are often given in some space environments [41-45], for example, in the Saturn's E-ring or G-ring, and the solar wind plasma etc. The data of dusty plasma in the Saturn's E-ring are listed in table 1 and in CGS system of units. The ion with mass $30.06\times10^{-24}$g is assumed $H_2O^+$, one of the water group ions [41], such as $O^+$, $OH^+$, $H_2O^+$, or $H_3O^+$. For convenience, the scattering factor is set $\ln\Lambda = 1$, which does not affect the function property of $(\partial\mathbf{U}/\partial t)_{t=0}$. Here we can take these data only as an example of the dusty plasma to numerically analyze the effects of each plasma component on the slowing down of the charged particle beam passing through the $\kappa$-distributed dusty plasma.

Table 1. Data of the dusty plasma in Saturn's E-ring

| Mass (g) | Density (cm$^{-3}$) | Temperature (K) | Charge (esu) |
|---|---|---|---|
| $m_e$=9.11×10$^{-28}$ | $n_e$=70 | $T_e$=4.642×10$^5$ | $q_e$= –4.8×10$^{-10}$ |
| $m_i$=30.06×10$^{-24}$ | $n_i$=1.0×10$^2$ | $T_i$= $T_e$/2 | $q_i$=4.8×10$^{-10}$ |
| $m_d$=10$^{-10}$ | $n_d$=0.1 | $T_d$= $T_e$/10 | $q_d$=300 $q_e$ |

In the following, as an example, we give the numerical analyses for the slowing down properties when an electron beam is passing through the $\kappa$-distributed dusty plasma, We draw the functional relationships of the electron term $(\partial\mathbf{U}/\partial t)_e \sim \mathbf{U}$, the ion term $(\partial\mathbf{U}/\partial t)_i \sim \mathbf{U}$ and the dust particle term $(\partial\mathbf{U}/\partial t)_d \sim \mathbf{U}$ by substituting the parameters in Table 1 into Eqs.(25), (26) and (27), respectively. In the three terms Eqs.(25)~(27), dependences of each term on the $\kappa$-parameter are in different ranges of the mean velocity U of the electron beam. What is more, we also draw the slowing down time $\tau$ as a function of **U**, on the basis of Eq.(21).

The effect of a certain kind of plasma component on the slowing down of a test particle beam becomes significant only if the mean speed U of the beam is *less than or equal to* the thermal speed of the plasma component [2]. Therefore, the order of magnitude of the horizontal axis coordinate (i.e. the mean speed U of the test particle beam) should be selected as that of thermal speed of the plasma component. In our case, the thermal speeds of the three plasma components are respectively, $V_{Te} \approx 2.7\times10^8$ cm/s for the electrons, $V_{Ti} \approx 1.0\times10^6$ cm/s for ions, and $V_{Td} \approx 0.25$ cm/s for dust particles, and so in the following figures, the order of magnitude of the horizontal axis coordinate (i.e. the mean speed U of the particle beam) is selected as $10^9$ cm/s for the electron term in Figs.(a), $10^7$ cm/s for the ion term in Figs.(b), and 1 cm/s for the dust term in Figs.(c).

## 4.1 *The slowing down of an electron beam in the $\kappa$-distributed dusty plasma*

Contributions of the electron term, the ion term and the dust particle term in Eq.(24) to the slowing down of an electron beam are numerically analyzed and are shown in Fig.1(a), Fig.1(b), and Fig.1(c), respectively. In Fig.1(d), we give the contributions of the dusty plasma (sum of the three terms in Eq.(24)) to the slowing down if the electrons, the ions and the dust particles have the same $\kappa$-parameter. All the four figures are drawing in a coordinate system, where the vertical axis is the deceleration factor and the horizontal axis is the mean velocity **U** of the electron beam. In Fig.1(e), we show the dependence of the slowing down time in Eq.(21) on the mean velocity of the electron beam. In all the five figures, the $\kappa$-parameters are chosen four different values, where $\kappa=\infty$ corresponds to the dusty plasma with a Maxwellian distribution and the other three correspond to the dusty plasma with $\kappa$-distribution.

We find the following results:

(1). In Figs.1(a) - 1(d), we show that the slowing down of the electron beam in the $\kappa$-distributed dusty plasma is stronger than that in a Maxwell-distributed dusty plasma, and the smaller the $\kappa$-parameter (the farther away from the Maxwellian distribution), the stronger the slowing down. And such differences will gradually decrease as increase of the mean velocity of the electron beam.

(2). In Figs.1(a)-1(c), we show that, the three terms in Eq.(24), $(\partial \mathbf{U}/\partial t)_e$, $(\partial \mathbf{U}/\partial t)_i$ and $(\partial \mathbf{U}/\partial t)_d$, compared with each other, the slowing down of the electron term on an electron beam is the weakest, the ion term is the second, and the dust particle term is the strongest. In Fig.1(c) and Fig.1(d), we see that the dust particle component plays a dominant role in the slowing down of an electron beam in the plasma.

(3). Fig.1(e) shows that there is little difference for the effect of the four different $\kappa$-parameters on the slowing down time of an electron beam in the dusty plasma. For the slowing down time of the electron beam, the $\kappa$-distributed dust plasma is almost the same as the Maxwell distributed plasma in this plasma example.

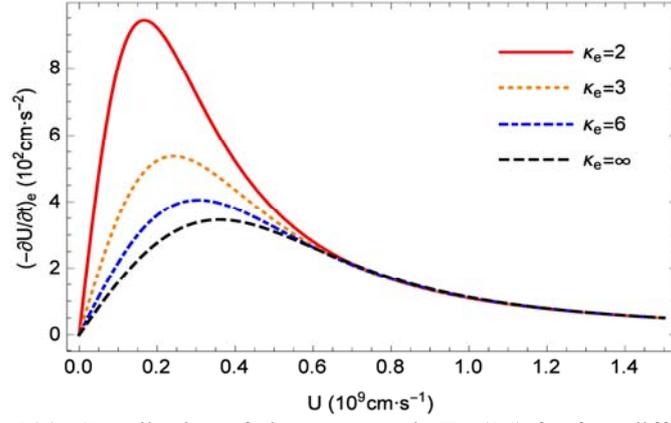

Fig.1(a). Contribution of electron term in Eq.(24) for four different $\kappa$ values to the slowing down of an electron beam.

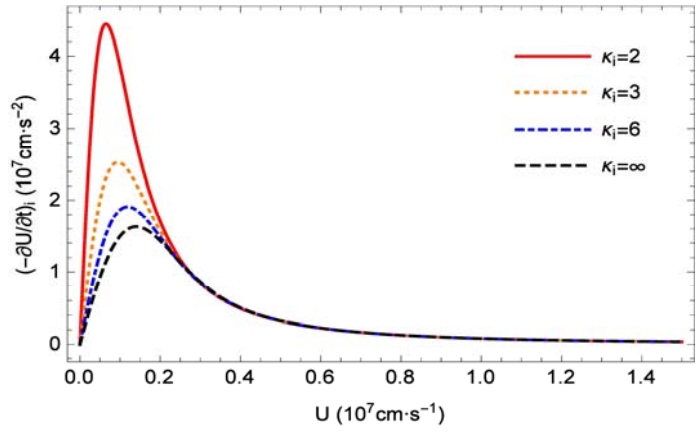

Fig.1(b). Contribution of ion term in Eq.(24) for four different $\kappa$ values to the slowing down of an electron beam.

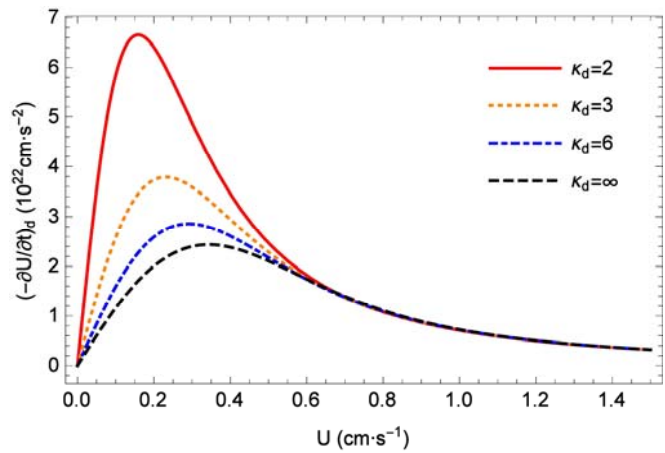

Fig.1(c). Contribution of dust particle term in Eq.(24) for four different $\kappa$ values to the slowing down of an electron beam.

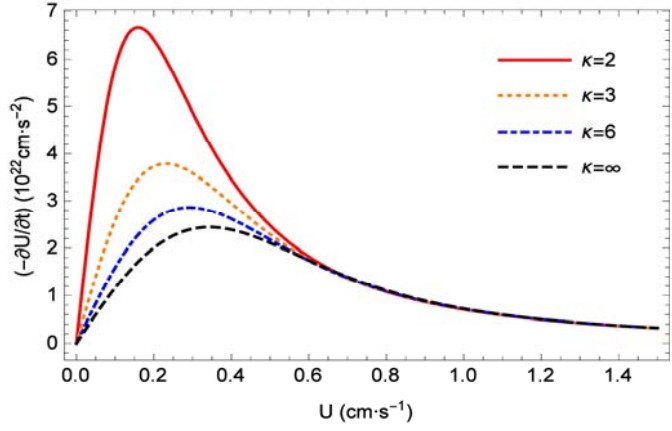

Fig.1(d). Total contribution of terms in Eq.(24) for four different $\kappa$ values to the slowing down of an electron beam if the $\kappa$-parameters of three plasma components are assumed to be the same.

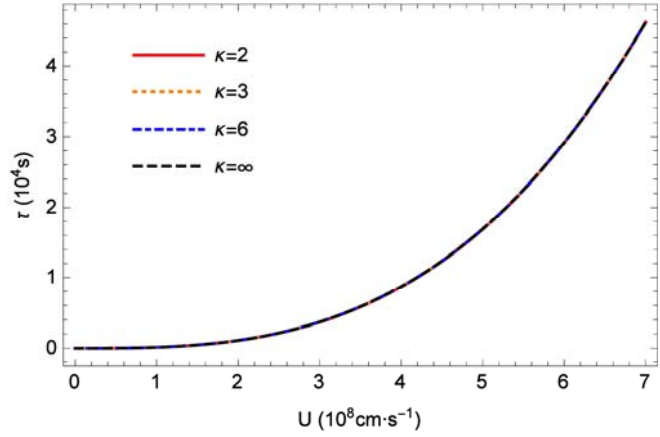

Fig.1(e). The slowing down time in Eq.(21) for four different $\kappa$ values if the $\kappa$-parameters of three components of the dusty plasma are assumed to be the same.

### 4.2 *The roles of charge and mass of a dust particle in the slowing down*

In this section, we give the numerical analyses for the effect of different charges and masses of the dust particles in the $\kappa$-distributed dusty plasma on the slowing down. As an example, here we consider the particle beam to be a proton beam and we only calculate the dust particle term in Eq.(24) because the dust particles play a dominant role in the slowing down of a particle beam. The results are shown in Fig.2 (a) and Fig.2 (b) respectively. In the two figures, the vertical axis is the deceleration factor of dust particle term and the horizontal axis is the mean velocity U of the proton beam.

In Fig.2(a), we show the contributions of dust particle term in Eq.(24) to the slowing down of a proton beam for four different mass $m_d$ of a dust particle in the $\kappa$-distributed dusty plasma with $\kappa=3$. It is shown that the slowing down of the proton beam depends obviously on mass of the dust particle in the dusty plasma only if the mean velocity of the beam is less than about $10^2$ cm s$^{-1}$, and the greater the mass of the dust particle, the stronger the deceleration effect of the proton beam.

In Fig.2(b), we show contributions of the dust particle term in Eq.(24) to the slowing down of a proton beam for four different charges $q_d$ of a dust particle in the $\kappa$-distributed dusty plasma with $\kappa=3$. It is shown that the slowing down of the proton beam depends obviously on the charge of the dust particle in the dusty plasma in all the range of the mean velocity of the beam, and the greater the charge of the dust particle, the stronger the deceleration effect of the particle beam.

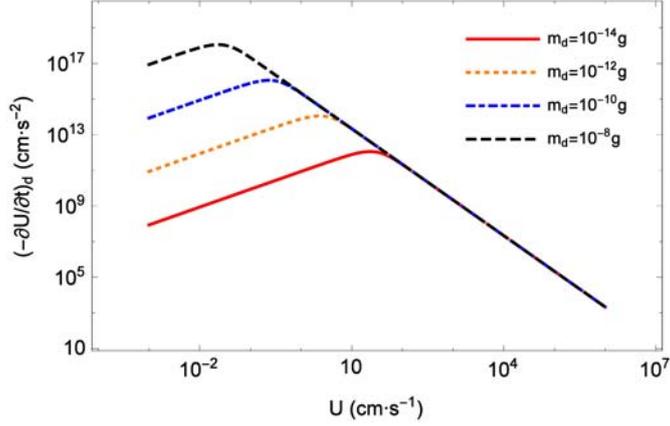

Fig. 2(a). Contribution of dust particle term in Eq.(24) to the slowing down of a proton beam for four different mass $m_d$ of dust particle in the $\kappa$-distributed dusty plasma with $\kappa=3$.

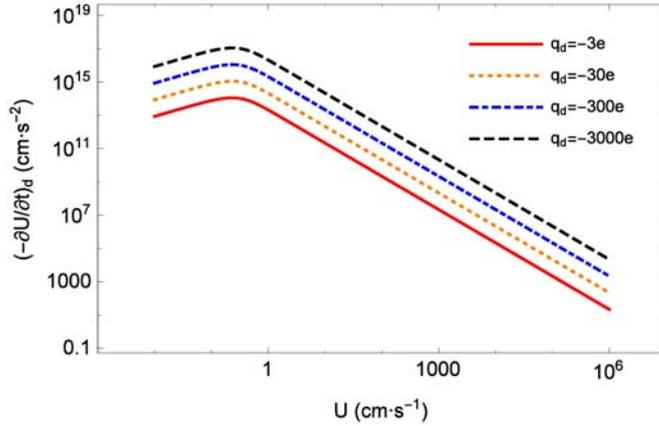

Fig. 2(b). Contribution of dust particle term in Eq.(24) to the slowing down of a proton beam for four different charge $q_d$ of dust particle in the $\kappa$-distributed dusty plasma with $\kappa=3$.

## 5 Conclusions

In conclusion, we have studied the FP collision theory for the slowing down phenomenon of charged particle passing through the dusty plasmas with power-law $\kappa$-distributions. The three plasma components, electrons, ions and dust particles, can have different $\kappa$-parameter. Using the FP collision theory and the function expressions

Eq.(7) of the power-law $\kappa$-distributions, we derived the deceleration factor (the velocity moment equation) of a test particle moving in the $\kappa$-distributed dusty plasma, Eq.(18), which contains a sum of the contributions of the three terms for the electrons, the ions and the dust particles in the plasma. We showed that the deceleration factor can be expressed by a hyper-geometric function for the $\kappa$-parameters as well as the mean velocity of the test particle. Furthermore, we obtained the slowing down time of the test particle passing through the $\kappa$-distributed dust plasma, given by Eq.(21). We see that all the equations can recover to the standard forms in the Maxwell-distributed plasmas when we take the $\kappa$-parameters to be infinite.

In order to more accurately show the roles of the different $\kappa$-parameters in the slowing down phenomena, we have made numerical analyses on the deceleration factors in Eq.(24) as a function of the mean velocity. The electron beam is chosen as an example of the charged particle beam in the numerical analyses. The dusty plasma example for the numerical analyses is from the data of the Saturn's E-ring dusty plasma. The $\kappa$-parameters are chosen as four different values: 2, 3, 6, and $\infty$, where the $\kappa$-parameter to be $\infty$ corresponds to the case of the dusty plasma with a Maxwellian velocity distribution. We also made numerical analyses on the slowing down time for the four different $\kappa$-parameters.

We show that the slowing down effect of the $\kappa$-distributed dusty plasma on an electron beam is stronger than that of the Maxwell-distributed dusty plasma, and the smaller the $\kappa$-parameter (the father away from the Maxwellian distribution), the stronger the deceleration effect. And such differences will gradually decrease as the increase of the mean velocity of the electron beam. In the contributions of the dusty plasma to the slowing down, the dust particles play a dominant role.

However, we find that there is little difference for different $\kappa$-parameters in the slowing down time of an electron beam in the dusty plasma. It is almost the same for the $\kappa$-distributed plasma and the Maxwell-distributed plasma in this dusty plasma example.

We also show that dust particle component plays a dominant role in the slowing down of a charged particle beam in the dusty plasma, and the slowing down not only depend strongly on the $\kappa$-parameters but also depend on the dust charge and dust mass. The greater charge and the greater mass the dust grain has, the stronger the deceleration effects of the beam displays in the dusty plasma.

**Acknowledgments**

This work was supported by the National Natural Science Foundation of China under grant No.11775156.

**Appendix**

On the right-hand side in (13), we use symbol $A_\alpha \equiv m_\alpha/[(2\kappa_\alpha-3)kT_\alpha]$, so that

$$h_{\kappa,\alpha}(\mathbf{v}) = \frac{\pi m_T}{\mu_\alpha} \frac{kT_\alpha}{m_\alpha \mathrm{v}} \left(2 - \frac{3}{\kappa_\alpha}\right) B_{\kappa,\alpha} \left\{ \int_0^\infty dw \left[1 + A_\alpha (w-\mathrm{v})^2\right]^{-\kappa_\alpha} \right.$$
$$\left. - \int_0^\infty dw \left[1 + A_\alpha (w+\mathrm{v})^2\right]^{-\kappa_\alpha} \right\}. \tag{A.1}$$

If we take the substitution $y=\sqrt{A_\alpha}(w\text{-v})$ in the first integral in Eq.(A.1), we have that

$$\int_0^\infty dw\left[1+A_\alpha(w\text{-v})^2\right]^{-\kappa_\alpha}=\frac{1}{\sqrt{A_\alpha}}\left[2\int_0^{\sqrt{A_\alpha}\text{v}}dy(1+y^2)^{-\kappa_\alpha}+\int_{\sqrt{A_\alpha}\text{v}}^\infty dy(1+y^2)^{-\kappa_\alpha}\right], \quad (A.2)$$

and if we take the substitution $y=\sqrt{A_\alpha}(w\text{+v})$ in the second integral in Eq.(A.1), we have that

$$\int_0^\infty dw\left[1+A_\alpha(w+\text{v})^2\right]^{-\kappa_\alpha}=\frac{1}{\sqrt{A_\alpha}}\int_{\sqrt{A_\alpha}\text{v}}^\infty dy(1+y^2)^{-\kappa_\alpha}. \quad (A.3)$$

Substituting (A.2) and (A.3) into (A.1), we obtain the equation (14).